\documentclass[aps,prd,twocolumn,superscriptaddress,amsmath,amssymb]{revtex4}
\usepackage{graphicx} 
\usepackage{epstopdf} 
\usepackage{dcolumn} 
\usepackage{xcolor} 
\usepackage{amsmath,amssymb} 
\usepackage{hyperref} 
\usepackage[utf8]{inputenc} 
\usepackage[english]{babel} 
\usepackage[displaymath, mathlines]{lineno} 
\usepackage{blindtext} 
\usepackage{ifthen} 
\usepackage{orcidlink} 

\graphicspath{{figures/}} 

\newcommand{\lumi}{190}
\newcommand{\sqrts}{10.579}

\def\invfb    {\ensuremath{\mbox{\,fb}^{-1}}\xspace}
\def\invab    {\ensuremath{\mbox{\,ab}^{-1}}\xspace}

\def\epem     {\ensuremath{e^+e^-}\xspace}
\def\eetautau     {\ensuremath{e^+e^-\to\tau^+\tau^-}\xspace}
\def\taupipipi {\ensuremath{\tau^-\to\pi^-\pi^+\pi^-\nu_\tau}\xspace}

\newcommand{\mtau}{\ensuremath{m_{\tau}}}
\newcommand{\Mmin}{\ensuremath{M_{\rm min}}}
\newcommand{\range}{\ensuremath{[1.70, 1.85]}}

\newcommand{\NData}{583192}

\newcommand{\MCEstimatorBias}{0.40 \pm 0.03}

\newcommand{\SystBeamEnergy}{0.07}

\newcommand{\SystMomentumSF}{0.06}
\newcommand{\SystMisalign}{0.03}

\newcommand{\SystEstimator}{0.03}
\newcommand{\SystFitFunc}{0.02}

\newcommand{\SystMaterial}{0.03}
\newcommand{\SystISRFSR}{0.02}

\newcommand{\central}{1777.09}
\newcommand{\Stat}{0.08}
\newcommand{\SystTotal}{0.11}

\def\result {\ensuremath{\central \pm \Stat \pm \SystTotal }\xspace}

\newboolean{articletitles}
\setboolean{articletitles}{true} 

\RequirePackage{xspace}
\RequirePackage{relsize}

\def\belletwo{\mbox{Belle~II}\xspace}

\def\babar{\mbox{\slshape B\kern-0.1em{\smaller A}\kern-0.1em
    B\kern-0.1em{\smaller A\kern-0.2em R}}\xspace}

\mathchardef\PDelta="7101
\mathchardef\PXi="7104
\mathchardef\PLambda="7103
\mathchardef\PSigma="7106
\mathchardef\POmega="710A
\mathchardef\PUpsilon="7107
                 
\def\PB      {\ensuremath{B}\xspace}                 
                 
\def\PD      {\ensuremath{D}\xspace}

\def\PK      {\ensuremath{K}\xspace}

\def\Pe      {\ensuremath{e}\xspace}

\def\Pi      {\ensuremath{i}\xspace}

\def\epem       {\ensuremath{\Pe^+\Pe^-}\xspace}

\def\kaon  {\ensuremath{\PK}\xspace}
\def\Kbar  {\kern 0.2em\overline{\kern -0.2em \PK}{}\xspace}

\def\Kz    {\ensuremath{\kaon^0}\xspace}
\def\Kzb   {\ensuremath{\Kbar^0}\xspace}
\def\KzKzb {\ensuremath{\Kz \kern -0.16em \Kzb}\xspace}
\def\Kp    {\ensuremath{\kaon^+}\xspace}
\def\Km    {\ensuremath{\kaon^-}\xspace}

\def\KpKm  {\ensuremath{\Kp \kern -0.16em \Km}\xspace}

\def\D       {\ensuremath{\PD}\xspace}
\def\Dbar    {\kern 0.2em\overline{\kern -0.2em \PD}{}\xspace}

\def\Dz      {\ensuremath{\D^0}\xspace}
\def\Dzb     {\ensuremath{\Dbar^0}\xspace}
\def\DzDzb   {\ensuremath{\Dz {\kern -0.16em \Dzb}}\xspace}
\def\Dp      {\ensuremath{\D^+}\xspace}
\def\Dm      {\ensuremath{\D^-}\xspace}

\def\DpDm    {\ensuremath{\Dp {\kern -0.16em \Dm}}\xspace}

\def\Bbar    {\ensuremath{\kern 0.18em\overline{\kern -0.18em \PB}{}}\xspace}

\def\Y#1S{\ensuremath{\PUpsilon{(#1S)}}\xspace}

\def\Lbar {\ensuremath{\kern 0.1em\overline{\kern -0.1em\PLambda}}\xspace}

\def\to                 {\ensuremath{\rightarrow}\xspace}

\newcommand{\tev}{\ensuremath{\mathrm{\,Te\kern -0.1em V}}\xspace}
\newcommand{\gev}{\ensuremath{\mathrm{\,Ge\kern -0.1em V}}\xspace}
\newcommand{\mev}{\ensuremath{\mathrm{\,Me\kern -0.1em V}}\xspace}
\newcommand{\kev}{\ensuremath{\mathrm{\,ke\kern -0.1em V}}\xspace}
\newcommand{\ev}{\ensuremath{\mathrm{\,e\kern -0.1em V}}\xspace}
\newcommand{\gevc}{\ensuremath{{\mathrm{\,Ge\kern -0.1em V\!/}c}}\xspace}
\newcommand{\mevc}{\ensuremath{{\mathrm{\,Me\kern -0.1em V\!/}c}}\xspace}
\newcommand{\gevcc}{\ensuremath{{\mathrm{\,Ge\kern -0.1em V\!/}c^2}}\xspace}
\newcommand{\gevgevcccc}{\ensuremath{{\mathrm{\,Ge\kern -0.1em V^2\!/}c^4}}\xspace}
\newcommand{\mevcc}{\ensuremath{{\mathrm{\,Me\kern -0.1em V\!/}c^2}}\xspace}

\def\invfb   {\ensuremath{\mbox{\,fb}^{-1}}\xspace}

\def\invab   {\ensuremath{\mbox{\,ab}^{-1}}\xspace}

\def\rad{\ensuremath{\rm \,rad}\xspace}

\def\gsim{{~\raise.15em\hbox{$>$}\kern-.85em
          \lower.35em\hbox{$\sim$}~}\xspace}
\def\lsim{{~\raise.15em\hbox{$<$}\kern-.85em
          \lower.35em\hbox{$\sim$}~}\xspace}

\begin{document}

\title{Measurement of the $\tau$-lepton mass with the Belle~II experiment

}
  \author{I.~Adachi\,\orcidlink{0000-0003-2287-0173}} 
  \author{K.~Adamczyk\,\orcidlink{0000-0001-6208-0876}} 
  \author{L.~Aggarwal\,\orcidlink{0000-0002-0909-7537}} 
  \author{H.~Ahmed\,\orcidlink{0000-0003-3976-7498}} 
  \author{H.~Aihara\,\orcidlink{0000-0002-1907-5964}} 
  \author{N.~Akopov\,\orcidlink{0000-0002-4425-2096}} 
  \author{A.~Aloisio\,\orcidlink{0000-0002-3883-6693}} 
  \author{N.~Anh~Ky\,\orcidlink{0000-0003-0471-197X}} 
  \author{D.~M.~Asner\,\orcidlink{0000-0002-1586-5790}} 
  \author{H.~Atmacan\,\orcidlink{0000-0003-2435-501X}} 
% ...
\collaboration{The Belle II Collaboration}

\begin{abstract}

We present a measurement of the $\tau$-lepton mass using a sample of about 175 million $e^+e^- \to \tau^+\tau^-$ events collected with the Belle II detector at the SuperKEKB $e^+e^-$ collider at a center-of-mass energy of $10.579\,\mathrm{Ge\kern -0.1em V}$. This sample corresponds to an integrated luminosity of $190\,\mathrm{fb^{-1}}$. We use the kinematic edge of the $\tau $ pseudomass distribution in the decay ${\tau^-\to\pi^-\pi^+\pi^-\nu_\tau}$ and measure the $\tau$ mass to be $1777.09 \pm 0.08 \pm 0.11 \,\mathrm{Me\kern -0.1em V\!/c^2}$, where the first uncertainty is statistical and the second systematic. This result is the most precise to date.

\end{abstract}

\maketitle

\label{sec:intro}

The $\tau$-lepton mass, \mtau, is one of the fundamental parameters of the standard model. 
An experimental determination of \mtau{} with the lowest possible uncertainty has important consequences for tests of lepton-flavor-universality between $\tau$ and lighter leptons where the $\tau$ mass enters to the fifth power~\cite{Amhis:2019ckw} . 
Precise knowledge of \mtau{} is also important for the predictions of leptonic and hadronic branching fractions of the $\tau$~\cite{Tsai:1971vv} and the determination of the strong-interaction coupling $\alpha_s$ at the $\tau$-mass scale~\cite{Davier:2005xq,Davier_2014}. 

The most precise measurements of the $\tau$ mass currently available are reported in Table~\ref{tab:worldData}. The BES~\cite{BES:1995jmv}, KEDR~\cite{KEDR:2006xae}, and BES~III~\cite{BESIII:2014srs} collaborations measured the $\tau$ mass by analyzing the $\epem \to \tau^+\tau^-$ cross-section near the $\tau^+\tau^-$ production threshold. The Belle~\cite{Belle:2006qqw} and BaBar~\cite{BaBar:2009qmj} measurements use the pseudomass endpoint method~\cite{Albrecht:1992} at a center-of-mass energy near the $\Upsilon(4S)$ mass. 
The highest precision to date has been achieved by the BES~III collaboration.
While the statistical and systematic uncertainties for the BES~III measurement are of similar size, the precision of the Belle and BaBar measurements is limited by systematic uncertainties. For both experiments the largest systematic uncertainties arose from the knowledge of the beam energy and the momentum reconstruction of the $\tau$-decay products.

\begin{table}[h!]
    \centering
    \begin{tabular}{l l}
    \hline
    \hline
     Experiment   \hspace{2 cm} &  \mtau{} [\mevcc] \\ 
    \hline
    \vspace{-0.2 cm}\\
    BES~\cite{BES:1995jmv}        &   $1776.96^{\;\; +\,\; 0.18}_{\;\; -\,\; 0.21}\ ^{\;\;+\;\;0.25}_{\;\;-\;\;0.17}$\\
    KEDR~\cite{KEDR:2006xae}      &   $1776.80^{\;\;+\,\;0.25}_{\;\;-\,\;0.23\;\;}  \pm 0.15$  \\
    BES III~\cite{BESIII:2014srs} &   $1776.91 \pm 0.12^{\;\;+\;\;0.10}_{\;\;-\;\;0.13}$ \\
    Belle~\cite{Belle:2006qqw}    &   $1776.61 \pm 0.13 \pm 0.35$   \\
    BaBar~\cite{BaBar:2009qmj}    &   $1776.68 \pm 0.12 \pm 0.41$ \\
    [0.1cm] \hline     \hline
    \end{tabular}
    \caption{Summary of most precise measurements of the $\tau$ mass to date.}
   \label{tab:worldData}
\end{table}

\begin{figure}[h!t]
    \centering
    \includegraphics[width=\linewidth]{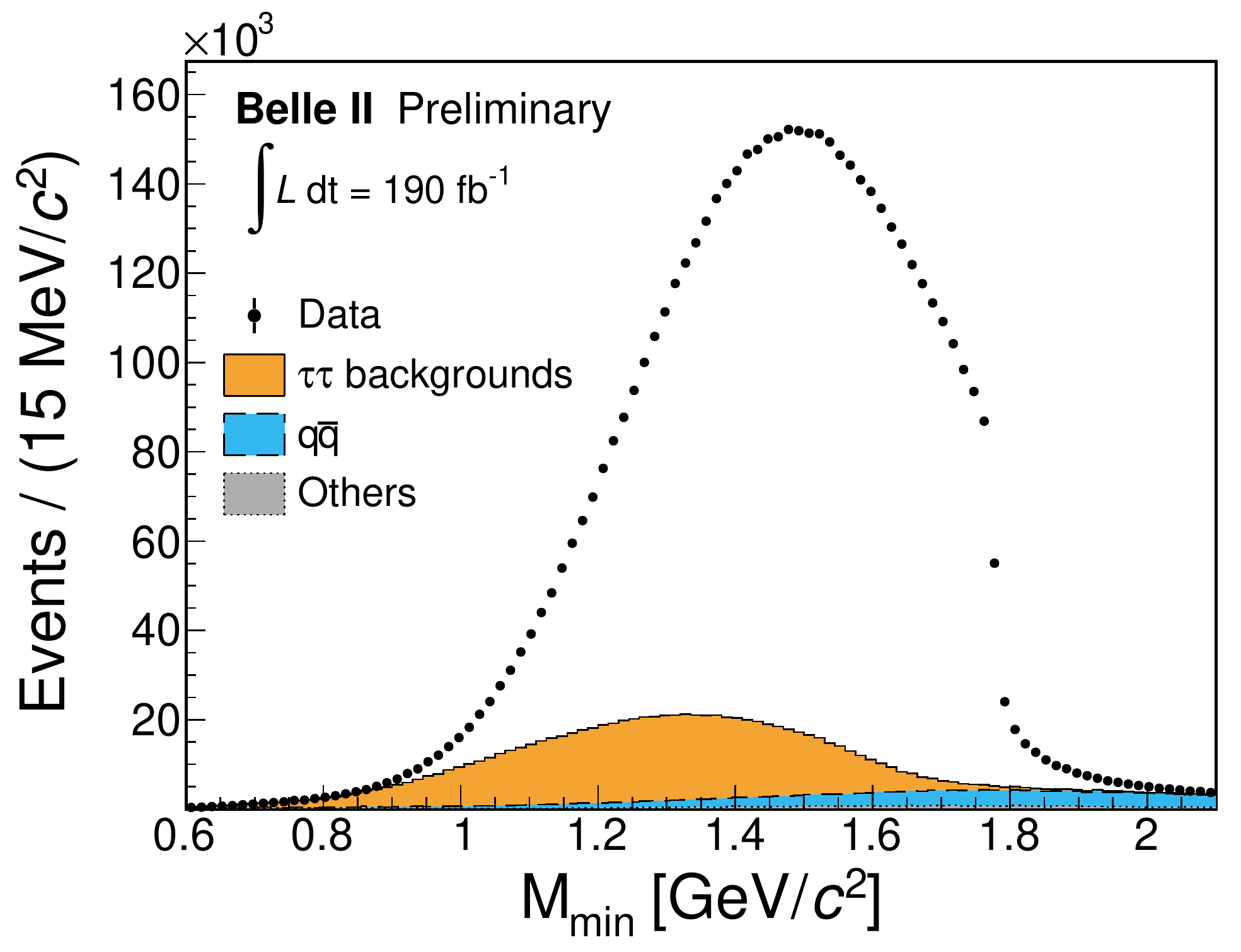}
    \caption{
    Spectrum of \Mmin{} in experimental data (dots), along with simulated background contributions from $e^+e^- \to \tau^+\tau^-$ events with decays other than \taupipipi{} (orange area with solid line), $e^+e^- \to q\bar{q}$ events (blue area with dashed line), and other background sources (gray area with dotted line).
    }
    \label{fig:mmin}
\end{figure}

In this paper, we report a measurement of \mtau{} using a sample 
of about 175 million $e^+e^-\to\tau^+\tau^-$ events recorded with the \belletwo{} detector~\cite{Abe:2010gxa} at the asymmetric-energy \epem{} SuperKEKB collider~\cite{Akai:2018mbz}. 
The data, collected between March 2019 and July 2021 near the nominal center-of-mass energy of $\sqrt{s}=\sqrts$\gev, correspond to an integrated luminosity of \lumi\invfb. 
We determine \mtau{} from the hadronic decays \taupipipi{} using the pseudomass endpoint method. Charge-conjugate modes are implied throughout. Assuming zero mass for the neutrino, the $\tau$ mass is given by 
\begin{equation}
    \mtau =\sqrt{M_{3\pi}^2 + 2( E_\tau^* - E^*_{3\pi})(E^*_{3\pi} - p^*_{3\pi}\cos\alpha^*)} .
	\label{eq:Mtau}
\end{equation}
Here, and throughout the paper, quantities in the $\epem$ center-of-mass frame are indicated by the asterisk. The mass, energy, and momentum of the three-pion system are denoted by $M_{3\pi}$, $E^*_{3\pi}$, and $p^*_{3\pi}$, respectively. 
The energy of the $\tau$ is given by $E_\tau^*$; $\alpha^*$ is the angle between the momenta of the three-pion system and the neutrino. 
The energy $E_\tau^*$ is half of the \epem center-of-mass energy $\sqrt{s}/2$ up to corrections from initial state radiation (ISR) from the $e^\pm$ beams and final state radiation (FSR) from the $\tau$ and its decay products.  
The pseudomass \Mmin{} is defined by setting $\alpha^*$ equal to 0 and therefore minimizing Eq.~\ref{eq:Mtau} as
\begin{equation}
	\Mmin = \sqrt{M^{2}_{3\pi} + 2(\sqrt{s}/2 - E^*_{3\pi})(E^*_{3\pi} - p^*_{3\pi})} \leq \mtau.
	\label{eq:mmin}
\end{equation}   
In the absence of ISR and FSR, and assuming a perfect measurement of the four-momentum of the three-pion system, the \Mmin{} distribution extends up to \mtau{}, where it has a sharp edge. The momentum resolution of the detector and the energy loss through radiation smear the endpoint position and introduce a tail towards larger \Mmin{} values. However, as seen in Fig.~\ref{fig:mmin}, an edge remains in the observed \Mmin{} distribution and is used to measure the $\tau$ mass. 
One challenge is to precisely measure the inputs to Eq.~\ref{eq:mmin}, namely the \epem center-of-mass energy $\sqrt{s}$ and the momenta of the $\tau$-decay products.  Another challenge is to develop an empirical model  to describe the \Mmin{} distribution. Any inaccuracy in either directly impacts the $\tau$-mass determination.  

\section{The Belle II detector and simulation} 
\label{sec:BelleII}

The Belle II detector consists of several subdetectors arranged in a cylindrical structure around the $e^+e^-$ interaction point~\cite{Abe:2010gxa}. 
Charged-particle trajectories (tracks) are reconstructed by a two-layer silicon-pixel detector, surrounded by a four-layer double-sided silicon-strip detector and a central drift chamber (CDC). Only 15\% of the second pixel layer was installed when the data were collected. 
Outside the CDC, a time-of-propagation detector and an aerogel ring-imaging Cherenkov detector cover the barrel and forward endcap regions, respectively. 
The electromagnetic calorimeter (ECL), divided into the forward endcap, barrel, and backward endcap, fills the remaining volume inside a 1.5 T superconducting solenoid and is used to reconstruct photons and electrons.  
A $K_L^0$ and muon detection system is installed in the iron flux return of the solenoid.  
The $z$ axis of the laboratory frame is defined as the detector solenoid axis, with the positive direction along the electron beam. The polar angle $\theta$ and the transverse plane are defined relative to this axis.

Several processes contribute to the \eetautau{} sample as backgrounds, including $\epem \to q\bar{q}$ events, where $q$ indicates a $u$, $d$, $c$, or $s$ quark; 
$\epem\to \epem (\gamma)$ and $\mu^+\mu^-(\gamma)$ events;
$\epem \to \ell^+\ell^-\ell^+\ell^-$ events, where $\ell$ is a charged lepton; 
$\epem \to e^+e^-h^+h^-$ events, where $h$ indicates a pion, kaon, or proton; and 
$\epem \to e^+e^- nh$ events with $n>2$. 
We use simulated events to identify discriminating features effective to suppress these backgrounds. 
The \eetautau{} process is generated using the KKMC generator~\cite{Jadach:1999vf, Jadach:2000ir}. The $\tau$ decays are simulated by TAUOLA~\cite{Jadach:1990mz} and their FSR by PHOTOS~\cite{Barberio:1990ms}. 
We use KKMC to simulate $\mu^+\mu^-(\gamma)$ and $q\bar{q}$ production; PYTHIA~\cite{Sjostrand:2014zea} for the fragmentation of the $q\bar{q}$ pair; BabaYaga@NLO~\cite{Balossini:2006wc, Balossini:2008xr, CarloniCalame:2003yt, CarloniCalame:2001ny,CarloniCalame:2000pz} for $\epem \to \epem (\gamma)$ events; and AAFH~\cite{BERENDS1985421,BERENDS1985441,BERENDS1986285} and TREPS~\cite{Uehara:1996bgt} for the production of non-radiative final states $\ell^+\ell^-\ell^+\ell^-$ and $e^+e^-h^+h^-$. 
There is no generator to simulate the $\epem \to e^+e^- nh$ process. 
The \belletwo{} analysis software~\cite{Kuhr:2018lps, basf2-zenodo} uses the GEANT4~\cite{Agostinelli:2002hh} package to simulate the response of the detector to the passage of the particles.\section{Event selection}
\label{sec:selection}

In the \epem{} center-of-mass frame, the $\tau$ leptons are produced in opposite directions. 
Thus, the decay products of one $\tau$ are isolated from those of the other $\tau$, and they are contained in opposite hemispheres. 
The boundary between those hemispheres is the plane perpendicular to the $\tau$ flight direction, which is experimentally approximated by the thrust axis. The thrust axis is the unit vector $\hat{t}$ that maximizes the thrust value $\sum{|\hat{t}\cdot \vec{p}_i^{\, *}|}/\sum{|\vec{p}_i^{\, *}|}$, where $\vec{p}_i^{\, *}$ is the momentum of $i$th final-state particle in the \epem{}center-of-mass frame~\cite{Brandt:1964sa,Farhi:1977sg}. 

We define the \emph{signal} hemisphere as that containing three charged particles, which are assumed to originate from the \taupipipi{} decay, and require that the other hemisphere, named \emph{tag}, contains only one charged particle and up to one neutral pion. 
Thus, the tag side contains leptonic ($\tau^+\to e^+ \nu_e \bar\nu_\tau$ and $\tau^+\to \mu^+ \nu_\mu \bar\nu_\tau$) and hadronic (predominantly $\tau^+\to h^+ \bar\nu_\tau$ and $\tau^+\to h^+ \pi^0 \bar\nu_\tau$ ) $\tau$ decays.

We select $\tau$-pair candidates by requiring the event to contain exactly four charged particles with zero total charge, each having a trajectory displaced from the average interaction point by less than 3~cm along the $z$ axis and less than 1~cm in the transverse plane to reduce the contamination of tracks originated from beam-background interactions. No particle-identification requirements are imposed on any of the charged particles. 
The momenta of charged particles are scaled with factors that range from $0.99660$ to $1.00077$ depending on the charge and $\cos\theta$ to correct for imperfections in the magnetic-field description used in the event reconstruction, misalignment of the detector, and material mismodeling. The correction factors are evaluated by measuring the mass-peak position of high-yield samples of $D^{0}\rightarrow K^{-} \pi^{+}$ decays reconstructed in data and comparing them to the known  value~\cite{ParticleDataGroup:2022pth}. 

Neutral pions are identified as photon pairs with masses between 115\mevcc{} and 152\mevcc.
Those photons are identified from ECL energy deposits (clusters), reconstructed within the CDC acceptance, $17^\circ < \theta < 150^\circ$, to ensure they are not matched to any charged particle. 
Depending on whether the photons are reconstructed in the forward, barrel, or backward region of the detector, requirements are different. Photon-energy thresholds ranging from 60\mev to 600\mev suppress the beam-induced backgrounds, which are larger in the endcaps compared to the barrel region. Requirements on the cosine of the angle between the momenta of the two photons and on the momentum of the reconstructed neutral pion reduce the combinatorial background from low-energy photons.

The online event selection, \emph{trigger}, is based on the energy deposits and their topologies in the ECL. The trigger efficiency is driven by the requirements of at least three clusters with a topology inconsistent with a Bhabha event and one of the clusters having an energy larger than 300 MeV. The trigger efficiency in the experimental data is approximately $92\%$. 

Background from $q\bar{q}$ production is suppressed by rejecting events containing neutral pions in the signal hemisphere or more than one neutral pion in the tag hemisphere. Events with photons of energy greater than 0.2\gev that are not used in neutral-pion reconstruction are also rejected. 
In order to suppress $\epem(\gamma)$, $\mu^+\mu^-(\gamma)$, $\ell^+\ell^-\ell^+\ell^-$, and $e^+e^-h^+h^-$ events, which are characterized by low-momentum charged particles, we rank the three charged particles in the signal hemisphere in decreasing order of transverse momenta and ensure that their values exceed 0.6, 0.2, and 0.1\gevc.
We further reject events from $q\bar{q}$, $\epem(\gamma)$, and $e^+e^- (nh)$ processes by restricting the thrust value and the visible center-of-mass energy of the event, $E^*_{\rm vis}$, that is the sum of all reconstructed particles' energies, to be within ranges of [0.87,0.97] and [2.5, 9.0]\gev, respectively.  
We remove most of the remaining background with requirements on the magnitude of the missing momentum, $0.05 < p^*_{\rm miss} < 3.5$\gevc, on its polar angle,  $0.5 < \theta^*_{\rm miss}< 2.7$\rad, and on the square of the missing mass, $0 < M^2_{\rm miss} < 54 \gevgevcccc$. The missing momentum is the difference between the momenta of the initial \epem{} and that of all reconstructed particles in the event, 
while the square of missing mass is defined as $M^2_{\rm miss} = (\sqrt{s}-E^*_{\rm vis})^2-(p^*_{\rm miss})^2$.

After all requirements, we observe \NData{} events in the experimental data in the \Mmin{} range \range\gevcc{}.  The signal-reconstruction efficiency in this region is $2.3\%$, and the purity of the sample is $90\%$. 
Among the signal events, around $56\%$ are lepton tagged and the remaining events are hadron tagged. 
The dominant backgrounds are from the $e^+e^- \to q\bar{q}$ processes ($6.4\%$), followed by $e^+e^- \to \tau^+\tau^- $ with decays other than \taupipipi{} in the signal hemisphere ($2.0\%$). \section{Beam energy}
\label{sec:BE}

The computation of \Mmin{} relies on the knowledge of the \epem{} center-of-mass energy. 
We exploit the fact that the collision energy is just slightly above the kinematic-production threshold for $B\bar{B}$  pairs and 
measure the $B$-meson energy, $E_B^*$, using fully reconstructed neutral and charged $B$-meson decays,
\begin{equation}
    E_B^* = \sqrt{m_B^2 + (p^*_B)^2} \approx m_B + \frac{1}{2m_B} (p^*_B)^2.
\end{equation}
Here $p_B^*$ and $m_B$ are the momentum and mass of the $B$ meson. 
A maximum-likelihood fit is performed to the $E_B^*$ distribution to determine its peak position.  
The collision energy $\sqrt{s}$ is obtained from $E_B^*$ after correcting for the effect of ISR and by accounting for the energy dependence of the $\epem \to B\bar{B}$ cross-section~\cite{BaBar:2008cmq, Belle:2021lzm} using the following procedure.

\begin{figure}
    \centering
    \includegraphics[width=\linewidth]{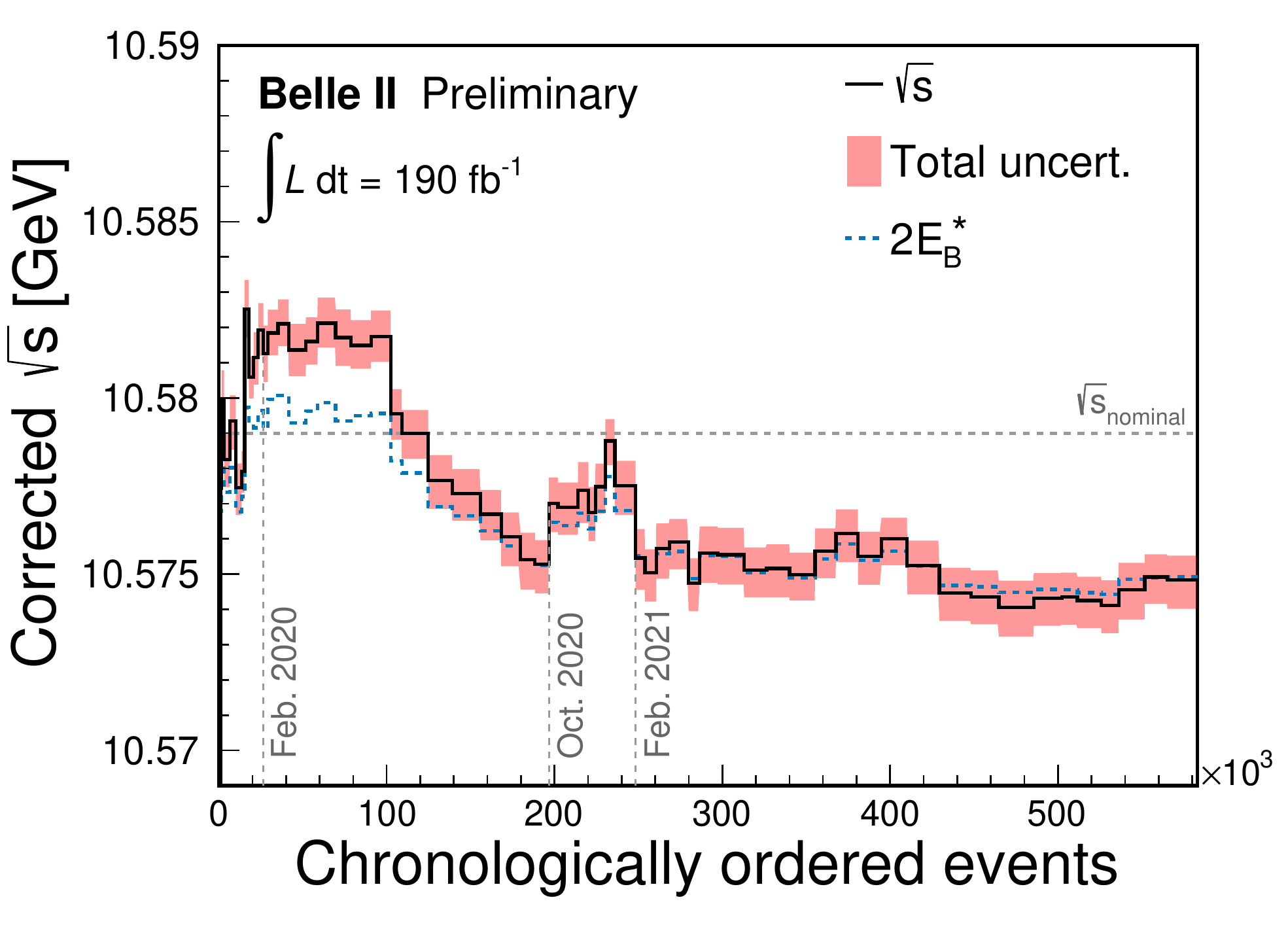}
    \caption{Corrected center-of-mass energy $\sqrt{s}$ (solid) and center-of-mass energy of $B \bar B$ pair $2E_B^*$ (dashed blue line) as functions of data-taking time, expressed in terms of chronologically ordered event numbers. The horizontal dashed line represents the nominal center-of-mass energy, $\sqrt{s}_{\rm nominal}=\sqrts\gev$, and the vertical dashed lines indicate the start of different data-taking periods. The 68.3\% confidence level band of $\sqrt{s}$ is displayed as shaded area.
    }
    \label{fig:EcmsCorr}
\end{figure}

We use a model where the center-of-mass energy of the colliding particles is described by a Gaussian of width $\sigma_{\sqrt{s}}$ about a mean value $\sqrt{s}$, where $\sqrt{s}$ and $\sigma_{\sqrt{s}}$ vary slowly during data-taking. The $\epem \to B\bar{B}$ cross-section is then given by
\begin{equation}
\frac{d^2 \sigma}{dx\, d\sqrt{s'}}  =  G(\sqrt{s'} - \sqrt{s}, \sigma_{\sqrt{s}}) \, W(s', x)\, \sigma_0 ( s' (1-x) ),
\label{eq:EcmsMain}
\end{equation}
where $\sqrt{s^\prime}$ is the event-by-event center-of-mass energy, $x$ is the fraction of energy carried by the ISR photon~\cite{Benayoun:1999hm}, and $G$ is the Gaussian distribution. 
The photon-emission probability is described by the function $W$, and $\sigma_0$ is the Born cross-section for $e^+e^- \to B\bar{B}$~\cite{Belle:2021lzm}. 
In terms of these quantities, the $B$-meson energy is given by $E^*_B = \frac{1}{2}\sqrt{s^\prime (1-x)}$. 
We use events simulated according to Eq.~\ref{eq:EcmsMain} to establish a mapping between the corrected quantities ($\sqrt{s},\sigma_{\sqrt{s}}$) and observed quantities ($E^*_B,\sigma_{E^*_B}$). 
The inverse of this mapping is used to obtain corrected values for the center-of-mass energy $\sqrt{s}$ as a function of data-taking time. The results are shown in Fig.~\ref{fig:EcmsCorr}. 
During the 2019 and early 2020 data-taking periods, the $\sqrt{s}$ value is around 2\mev above the nominal value $\sqrt{s}_{\rm nominal}=\sqrts$\gev. 
It then drifts to lower values where it stabilizes at around 6\mev below the nominal value by the middle of the 2021 data-taking period. 

\section{Method}
\label{sec:fitfunc}

To reduce experimenter's bias, we validate the method of the $\tau$-mass measurement and estimate the statistical and systematic uncertainties before looking at the central value of the result. 
The \Mmin{} distribution around the edge can be empirically described as a Heaviside step function multiplied by second-order polynomials, and convoluted with a double-Gaussian resolution function to account for the ISR, FSR, and detector resolution. We approximate this function with the following expression: 
\begin{eqnarray}
    \label{eq:pdf}   
     F(\Mmin) &=& 1-P_3\cdot \arctan \bigg(\frac{\Mmin - P_1}{P_2}\bigg) \nonumber \\
     &+& P_4 (\Mmin-P_1) + P_5 (\Mmin-P_1)^2 \;.
\end{eqnarray}
To determine the $\tau$ mass, we perform an unbinned maximum-likelihood fit~\cite{Verkerke:2003ir} to the \Mmin{} distribution in the range \range\gevcc using Eq.~\ref{eq:pdf}. 
The parameter $P_1$ determines the edge position and therefore is an estimator of the $\tau$ mass. The $P_2$ parameter modifies the slope of the threshold, while the rest of the parameters describe the shape away from the edge. 
Fits to simulated events in which the generated value of the $\tau$ mass is varied in the range [1772, 1782]\mevcc , show that on average $P_1$ overestimates the $\tau$ mass with a constant offset of $\MCEstimatorBias$\mevcc. 
This bias results from the empirical parametrization of the \Mmin{} distribution. 

\begin{figure}[h!t]
    \centering
    \includegraphics[width=\linewidth]{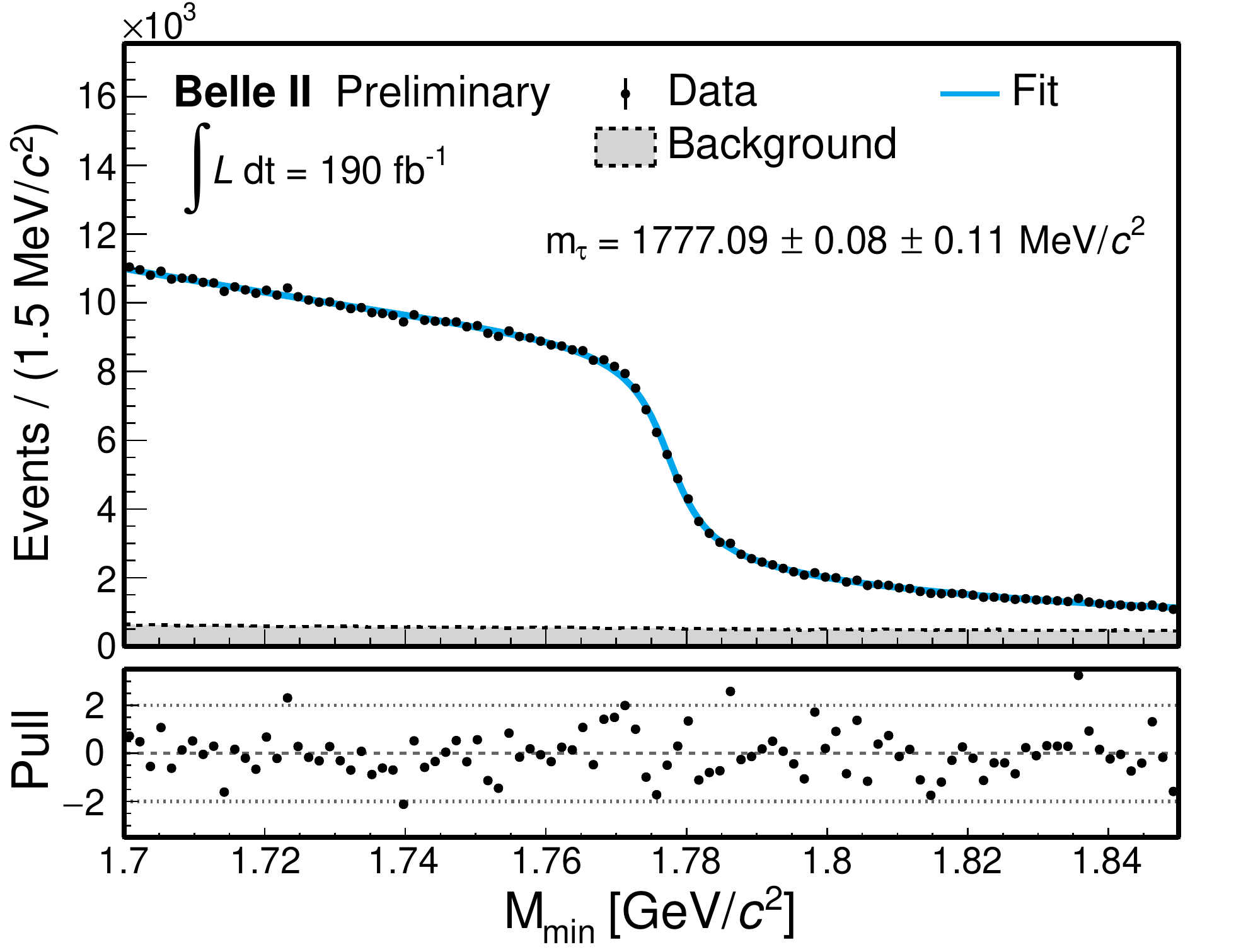}
    \caption{Top: Spectrum of \Mmin{} in experimental data (dots) with fit result (solid blue line) and background contribution (gray-filled area) overlaid.
    Bottom: Differences between data and fit result divided by the statistical uncertainties.  
    }
    \label{fig:fit}
\end{figure}

Figure~\ref{fig:fit} shows the \Mmin{} distribution in the range of \range\gevcc{} in data with the background predicted from simulation and the fit projection overlaid. While the \taupipipi{} events show a clear threshold, the background processes are featureless around the endpoint. 
Their contribution is described by the parameters $P_3$, $P_4$, and $P_5$. 
The observed value $P_1 = 1777.49 \pm \Stat$\mevcc is then corrected for the estimator bias to obtain the measured $\tau$ mass,
\begin{equation}
\mtau =  \central \pm \Stat \; \mevcc,
\label{eq:mtaucorr}
\end{equation}
where the uncertainty is the statistical uncertainty of the $P_1$ parameter. 
The statistical precision of the result is validated by generating simplified simulated experiments based on Poisson statistics, as well as by re-sampling the data based on bootstrapping techniques~\cite{Efron:Bootstrap}, and repeating the measurement on them.

\section{Systematic uncertainties and consistency checks }
\label{sec:sys}

The systematic uncertainties are grouped into categories associated with 
the knowledge of the colliding beams, 
the reconstruction of the charged particles, 
the fit model, and 
imperfections in the simulated data. 
Table~\ref{table:systsum} summarizes the sources that contribute to the total uncertainty. 
The largest uncertainty arises from the beam-energy correction, followed by the uncertainty on the charged-particle momentum. The various systematic uncertainties are added in quadrature, resulting in a total systematic uncertainty of $\SystTotal$\mevcc. 

\begin{table}
\begin{center}
\begin{tabular}{lr}
\hline
\hline
\vspace{-0.3 cm}\\
Source  & Uncertainty  \\
         &  [\mevcc]  \\
[0.1cm] \hline
\vspace{-0.3 cm}\\
Knowledge of the colliding beams: & \\
\quad Beam-energy correction                  & $\SystBeamEnergy$ \\
\quad Boost vector                            & $ < 0.01$  \\
[0.1cm] \hline
\vspace{-0.3 cm}\\
Reconstruction of charged particles: & \\
\quad Charged-particle momentum correction    & $\SystMomentumSF$ \\
\quad Detector misalignment                   & \SystMisalign    \\
[0.1cm] \hline
\vspace{-0.3 cm}\\
Fit model: & \\
\quad Estimator bias                          & \SystEstimator   \\ 
\quad Choice of the fit function              & \SystFitFunc     \\ 
\quad Mass dependence of the bias             & $ < 0.01$ \\ 
[0.1cm] \hline
\vspace{-0.3 cm}\\
Imperfections of the simulation: & \\
\quad Detector material density                & \SystMaterial   \\
\quad Modeling of ISR, FSR and $\tau$ decay   & \SystISRFSR          \\
\quad Neutral particle reconstruction efficiency  & $\leq$ 0.01  \\
\quad Momentum resolution                     & $ < 0.01$ \\ 
\quad Tracking efficiency correction          & $ < 0.01$  \\ 
\quad Trigger efficiency                      & $ < 0.01$  \\
\quad Background processes                    & $ < 0.01$ \\
[0.1cm] \hline 
\vspace{-0.25 cm}\\
Total                                         & $\SystTotal$  \\
[0.1cm] \hline
\hline
\end{tabular}
\end{center}
\caption{Summary of systematic uncertainties in the $\tau$-mass measurement.
}
\label{table:systsum}
\end{table} 

\subsection{Knowledge of the colliding beams}

The uncertainty on the $\sqrt{s}$ measurement, as indicated by the red band in Fig.~\ref{fig:EcmsCorr}, is on average around 0.75\mev and is dominated by systematic uncertainties. 
The estimation of $\sqrt{s}$ from the $B$-meson energy relies on the knowledge of the energy dependence of the $\epem\to B \bar{B}$ cross section, whose uncertainty is driven by the beam-energy uncertainty of the BaBar measurement~\cite{Belle:2021lzm, BaBar:2008cmq}. 
An additional uncertainty originates from the uncertainties in the average values of the charged ($0.26$\mevcc) and neutral ($0.20$\mevcc) $B$-meson masses~\cite{ParticleDataGroup:2022pth}.
Since the $\epem \to B \bar{B}$ sample is dominated by charged $B$ mesons, the weighted average of the two uncertainties results in $0.24$\mevcc, assuming the uncertainties of the $B^+$ and $B^0$ masses are fully correlated. 
In the experimental data the $B$-meson energy spread has an average value of 5.4\mev. A systematic uncertainty of 0.25\mev is assigned to the measurement of $\sigma_{E_B^*}$ based on simulation. 
The impact of the various systematic uncertainties on the $\tau$ mass is added in quadrature and results in $\SystBeamEnergy$\mevcc.

The computation of \Mmin{} relies on the knowledge of the boost vector of the center-of-mass frame. The boost vector is measured in experimental data using a dimuon sample. Its uncertainty is found to contribute negligibly to the systematic uncertainty on the $\tau$ mass.

\subsection{Reconstruction of charged particles}

\begin{figure}
    \centering
    \includegraphics[width=\linewidth]{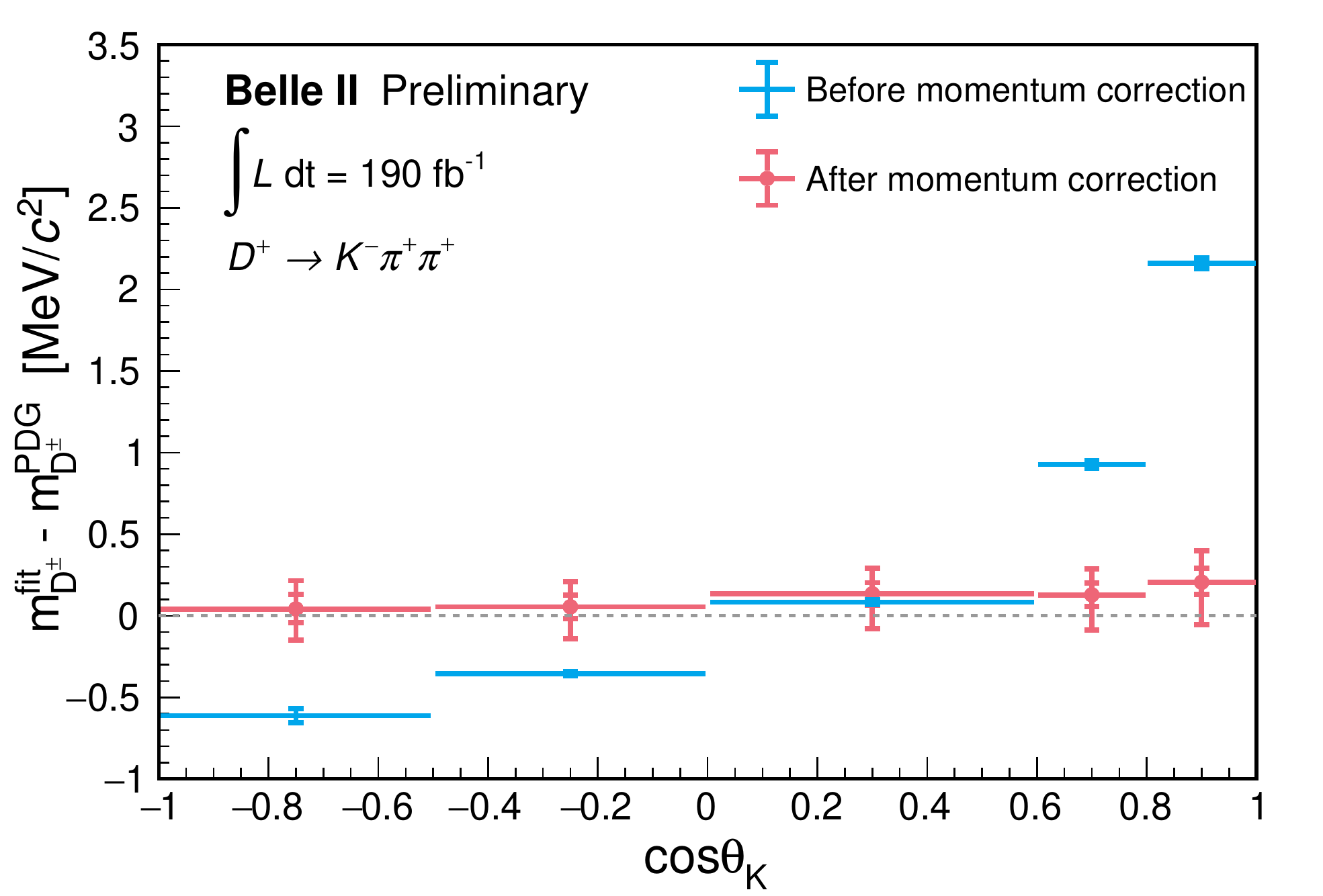}
    \caption{Deviation of the $D^{\pm}$ invariant-mass peak position from the known value before (blue) and after (red) momentum corrections as a function of the cosine of the kaon polar angle $\theta_K$. 
    The vertical error bars on the blue points indicate the statistical uncertainties in determining the peak position while the vertical error bars on the red points indicate the statistical and total uncertainties of the applied momentum corrections. 
    }
    \label{fig:dk2pi}
\end{figure}

Several sources of systematic uncertainty impact the determination of the charged-particle momentum corrections derived from the $D^0 \to K^-\pi^+$ sample. Different sources contribute to different $\cos\theta$ regions.
In particular, some residual dependence of the scale factors on the transverse momentum $p_{\rm T}$ is observed. This effect is included as a source of systematic uncertainty by measuring the variation in the scale factors after splitting the $D^0 \to K^-\pi^+$ sample in $p_{\rm T}$ at its median value of 1.3 \gevc.
The small deviation from the known value of the $D^{0}$ mass-peak observed in simulation is also included as systematic uncertainty.
Other important sources of uncertainties include the modeling of the $D^{0}$ mass peak, the uncertainty of the known $D^{0}$ mass, and a bias due to differences in the $\cos\theta$ distributions of the charged particles in the \taupipipi{} and $D^{0}\rightarrow K^{-} \pi^{+}$ samples.
As a consistency check, the scale factors are tested in $D^+\to K^{-}\pi^{+}\pi^{+}$, $D^0 \to K^{-}\pi^{+}\pi^{-}\pi^{+}$, and $J/\psi \to \mu^{-}\mu^{+}$ decays. The $D^+$,  $D^0$, and $J/\psi$ peak positions match the known values within the uncertainties, as shown in Fig.~\ref{fig:dk2pi} for the $D^\pm$. 
Assuming that the individual systematic effects are independent, the impact of the variations of the momentum correction on the $\tau$ mass are added in quadrature, resulting in \SystMomentumSF\mevcc.

The correction of tracking misalignment uses cosmic-ray tracks, and di-muon and hadronic collision-events~\cite{BelleIIPXD:2020rle}. 
To estimate the impact of a residual misalignment in the determination of \mtau{}, various misalignment configurations are used in the simulated data. 
The maximum deviation with respect to the nominal configuration, \SystMisalign\mevcc, is assigned as the systematic uncertainty due to the residual misalignment of the sub-detectors.

\subsection{Fit model}

The uncertainty of the estimator bias directly propagates to the precision of the $\tau$ mass, resulting in an uncertainty of \SystEstimator\mevcc. 
To test the independence of the estimator bias on the $\tau$ mass, an alternative assumption of a linear dependence is used, which results in the same bias.
Thus no additional systematic uncertainty is assigned.

The dependence of the result on the choice of the edge parametrization is investigated by repeating the measurement with alternative functions used previously by the Belle and BaBar collaborations~\cite{Belle:2006qqw,BaBar:2009qmj}. 
The largest deviation with respect to the main result is \SystFitFunc\mevcc and is assigned as a systematic uncertainty.

\subsection{Imperfections of the simulation}

We study possible simulation mismodelings that might lead to an incorrect estimation of the fit-bias correction. Differences between the properties of material used in the simulation and those in the detector may have an impact on the correction of the fit bias. Studies of the interaction of photons with the detector material indicate a deficit of around $10\%$ in the density of the beam-pipe in simulation. 
The impact of this deficit is tested by increasing by $10\%$ the beam-pipe density in the simulation of a signal sample corresponding to $4\invab$. 
The statistical uncertainty of the difference between the results using the nominal simulation and the simulation with the modified material density is \SystMaterial\mevcc, which is seen to be significantly larger than the actual difference between the two models. 
Hence, we assign the statistical precision of the difference as the uncertainty for this effect.

The modeling of ISR and FSR as well as the kinematic properties of the $\tau$-decay products may be different in simulated and experimental data. 
The simulated \Mmin{} distribution is weighted according to the observed differences between the experimental and simulated distributions in $p^*_{3\pi}$. 
The impact on the result is found to be \SystISRFSR\mevcc. 

Systematic uncertainties due to the simulation mismodeling of photon and neutral-pion reconstruction, transverse-momentum resolution, track-finding, trigger efficiencies, and background processes are found to be below or equal to 0.01\mevcc each.

\subsection{Consistency checks}

We check the stability of the result throughout various data-taking periods and observe no evidence for a time dependence. 
To exclude a potential dependence of the measured $\tau$ mass on the kinematic properties of the three-pion system or the $\tau$-decay products, we divide the data into sub-regions of various kinematic variables. Specifically, we use the cosine of the polar angle of the three-pion system and the individual pions, $M_{3\pi}$ and $p_{3\pi}$, and the momentum of the highest-momentum decay product. 
We obtain consistent results, indicating no significant unaccounted-for systematic effects.  
Finally, we explicitly test for a dependence of the measurement on the modeling of the $\tau$ decay. 
In the version of the TAUOLA program used for the simulation of $\tau$  decays~\cite{Jadach:1993hs} the modeling of the three-pion mass distribution in the \taupipipi{} channel is based on form factors from Ref.~\cite{Kuhn:1992nz}. 
As an alternative we use a sample simulated with form factors based on resonance chiral-Lagrangian currents for the hadronic $\tau$  decays~\cite{Banerjee:2021rtn, Shekhovtsova:2012ra, Chrzaszcz:2016fte, Nugent:2013hxa}. 
Using 6.6\invab of simulated samples, the fit to the generator-level \Mmin{} distributions of $\tau$ decays simulated with the two models show negligible variation in the resulting $P_1$ values. The $P_1$ values from fits to the reconstructed distributions are in agreement within $1.7\sigma$.  
Therefore no additional source of systematic uncertainty is considered.\section{Summary}
\label{sec:sum}

\begin{figure}
    \centering
    \includegraphics[width=\linewidth]{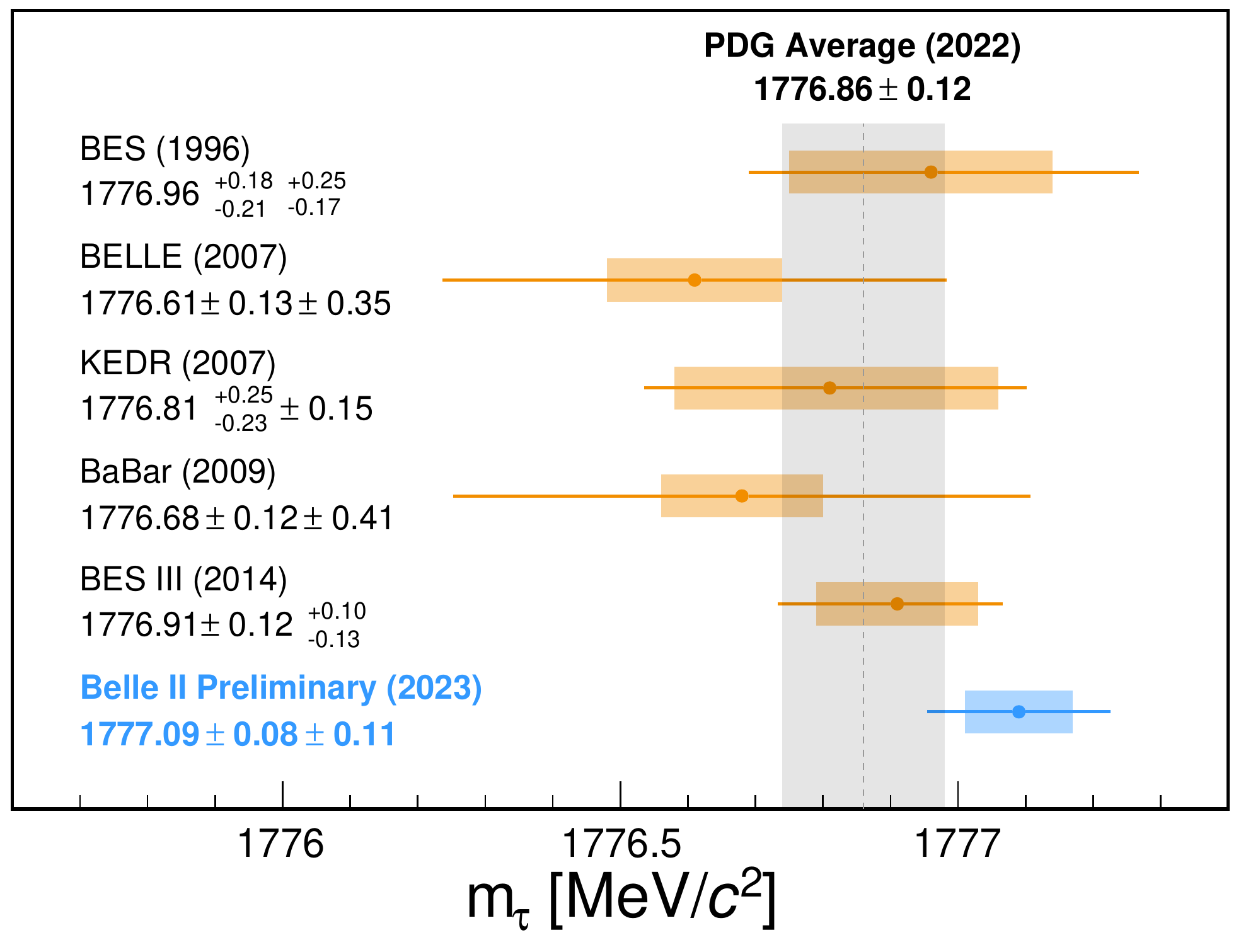}
    \caption{Summary of the most precise $\tau$-mass measurements~\cite{BES:1995jmv, Belle:2006qqw, KEDR:2006xae, BaBar:2009qmj, BESIII:2014srs} compared with the result of this work. The vertical gray band indicates the average value of previous measurements~\cite{ParticleDataGroup:2022pth}. 
    The inner bars represent the statistical uncertainties, while the outer bars indicate the total uncertainties. } 
    \label{fig:worldData}
\end{figure}

We measure the mass of the $\tau$ lepton to be
\begin{equation}
    \mtau =  \result{}\mevcc 
\end{equation}
using \eetautau{} data collected with the \belletwo{} detector at a center-of-mass energy of $\sqrt s = \sqrts$\gev and corresponding to an integrated luminosity of \lumi\invfb. 
The statistical uncertainty per unit sample size is smaller compared to the previous results~\cite{Belle:2006qqw, BaBar:2009qmj} owing to the improved event selection and momentum resolution of the \belletwo detector, which result in a steeper slope of the \Mmin{} distribution in the threshold region. 
The main sources of systematic uncertainty arise from the knowledge of the beam energy and from the uncertainty of the charged-particle momentum correction. 
As shown in Fig.~\ref{fig:worldData}, our result is consistent with previous measurements~\cite{BESIII:2014srs,BES:1995jmv,KEDR:2006xae,Belle:2006qqw,BaBar:2009qmj} and is the most precise to date.

This work, based on data collected using the Belle II detector, which was built and commissioned prior to March 2019, was supported by
Science Committee of the Republic of Armenia Grant No.~20TTCG-1C010;
Australian Research Council and research Grants
No.~DP200101792, 
No.~DP210101900, 
No.~DP210102831, 
No.~DE220100462, 
No.~LE210100098, 
and
No.~LE230100085; 
Austrian Federal Ministry of Education, Science and Research,
Austrian Science Fund
No.~P~31361-N36
and
No.~J4625-N,
and
Horizon 2020 ERC Starting Grant No.~947006 ``InterLeptons'';
Natural Sciences and Engineering Research Council of Canada, Compute Canada and CANARIE;
National Key R\&D Program of China under Contract No.~2022YFA1601903,
National Natural Science Foundation of China and research Grants
No.~11575017,
No.~11761141009,
No.~11705209,
No.~11975076,
No.~12135005,
No.~12150004,
No.~12161141008,
and
No.~12175041,
and Shandong Provincial Natural Science Foundation Project~ZR2022JQ02;
the Ministry of Education, Youth, and Sports of the Czech Republic under Contract No.~LTT17020 and
Charles University Grant No.~SVV 260448 and
the Czech Science Foundation Grant No.~22-18469S;
European Research Council, Seventh Framework PIEF-GA-2013-622527,
Horizon 2020 ERC-Advanced Grants No.~267104 and No.~884719,
Horizon 2020 ERC-Consolidator Grant No.~819127,
Horizon 2020 Marie Sklodowska-Curie Grant Agreement No.~700525 "NIOBE"
and
No.~101026516,
and
Horizon 2020 Marie Sklodowska-Curie RISE project JENNIFER2 Grant Agreement No.~822070 (European grants);
L'Institut National de Physique Nucl\'{e}aire et de Physique des Particules (IN2P3) du CNRS (France);
BMBF, DFG, HGF, MPG, and AvH Foundation (Germany);
Department of Atomic Energy under Project Identification No.~RTI 4002 and Department of Science and Technology (India);
Israel Science Foundation Grant No.~2476/17,
U.S.-Israel Binational Science Foundation Grant No.~2016113, and
Israel Ministry of Science Grant No.~3-16543;
Istituto Nazionale di Fisica Nucleare and the research grants BELLE2;
Japan Society for the Promotion of Science, Grant-in-Aid for Scientific Research Grants
No.~16H03968,
No.~16H03993,
No.~16H06492,
No.~16K05323,
No.~17H01133,
No.~17H05405,
No.~18K03621,
No.~18H03710,
No.~18H05226,
No.~19H00682, 
No.~22H00144,
No.~26220706,
and
No.~26400255,
the National Institute of Informatics, and Science Information NETwork 5 (SINET5), 
and
the Ministry of Education, Culture, Sports, Science, and Technology (MEXT) of Japan;  
National Research Foundation (NRF) of Korea Grants
No.~2016R1\-D1A1B\-02012900,
No.~2018R1\-A2B\-3003643,
No.~2018R1\-A6A1A\-06024970,
No.~2018R1\-D1A1B\-07047294,
No.~2019R1\-I1A3A\-01058933,
No.~2022R1\-A2C\-1003993,
and
No.~RS-2022-00197659,
Radiation Science Research Institute,
Foreign Large-size Research Facility Application Supporting project,
the Global Science Experimental Data Hub Center of the Korea Institute of Science and Technology Information
and
KREONET/GLORIAD;
Universiti Malaya RU grant, Akademi Sains Malaysia, and Ministry of Education Malaysia;
Frontiers of Science Program Contracts
No.~FOINS-296,
No.~CB-221329,
No.~CB-236394,
No.~CB-254409,
and
No.~CB-180023, and No.~SEP-CINVESTAV research Grant No.~237 (Mexico);
the Polish Ministry of Science and Higher Education and the National Science Center;
the Ministry of Science and Higher Education of the Russian Federation,
Agreement No.~14.W03.31.0026, and
the HSE University Basic Research Program, Moscow;
University of Tabuk research Grants
No.~S-0256-1438 and No.~S-0280-1439 (Saudi Arabia);
Slovenian Research Agency and research Grants
No.~J1-9124
and
No.~P1-0135;
Agencia Estatal de Investigacion, Spain
Grant No.~RYC2020-029875-I
and
Generalitat Valenciana, Spain
Grant No.~CIDEGENT/2018/020
Ministry of Science and Technology and research Grants
No.~MOST106-2112-M-002-005-MY3
and
No.~MOST107-2119-M-002-035-MY3,
and the Ministry of Education (Taiwan);
Thailand Center of Excellence in Physics;
TUBITAK ULAKBIM (Turkey);
National Research Foundation of Ukraine, project No.~2020.02/0257,
and
Ministry of Education and Science of Ukraine;
the U.S. National Science Foundation and research Grants
No.~PHY-1913789 
and
No.~PHY-2111604, 
and the U.S. Department of Energy and research Awards
No.~DE-AC06-76RLO1830, 
No.~DE-SC0007983, 
No.~DE-SC0009824, 
No.~DE-SC0009973, 
No.~DE-SC0010007, 
No.~DE-SC0010073, 
No.~DE-SC0010118, 
No.~DE-SC0010504, 
No.~DE-SC0011784, 
No.~DE-SC0012704, 
No.~DE-SC0019230, 
No.~DE-SC0021274, 
No.~DE-SC0022350, 
No.~DE-SC0023470; 
and
the Vietnam Academy of Science and Technology (VAST) under Grant No.~DL0000.05/21-23.

These acknowledgements are not to be interpreted as an endorsement of any statement made
by any of our institutes, funding agencies, governments, or their representatives.

We thank the SuperKEKB team for delivering high-luminosity collisions;
the KEK cryogenics group for the efficient operation of the detector solenoid magnet;
the KEK computer group and the NII for on-site computing support and SINET6 network support;
and the raw-data centers at BNL, DESY, GridKa, IN2P3, INFN, and the University of Victoria for offsite computing support.

\bibliography{references}

\end{document}